\pdfoutput=1

\documentclass[10pt,aps,pre,twocolumn,superscriptaddress,footinbib]{revtex4-1}

\pdfoutput=1
\usepackage{graphicx}
\usepackage{amssymb}
\usepackage{moresize}
\usepackage{amsmath}
\usepackage{bm}
\usepackage{txfonts}
\hbadness=\maxdimen
\vbadness=\maxdimen
\usepackage{titlesec}
\usepackage[usenames]{color}

\usepackage{placeins}

\def\be{\begin{equation}}
\def\ee{\end{equation}}
\def\bea{\begin{eqnarray}}
\def\eea{\end{eqnarray}}
\def\ba{\begin{array}}
\def\ea{\end{array}}

\begin{document}

\title{\large Compression stiffening of fibrous networks with stiff inclusions}


\author{Jordan Shivers}
\affiliation{Department of Chemical and Biomolecular Engineering, Rice University, Houston, TX 77005}
\affiliation{Center for Theoretical Biological Physics, Rice University, Houston, TX 77030}
\author{Jingchen Feng}
\affiliation{Center for Theoretical Biological Physics, Rice University, Houston, TX 77030}
\author{Anne S.\ G.\ van Oosten}
\affiliation{Institute for Medicine and Engineering, University of Pennsylvania, Philadelphia, PA 19104}
\author{Herbert Levine}
\affiliation{Center for Theoretical Biological Physics, Rice University, Houston, TX 77030}
\affiliation{Department of Physics and Department of Bioengineering, Northeastern University, Boston, MA 02115}
\author{Paul Janmey}
\affiliation{Institute for Medicine and Engineering, University of Pennsylvania, Philadelphia, PA 19104}
\author{Fred C.\ MacKintosh}
\affiliation{Department of Chemical and Biomolecular Engineering, Rice University, Houston, TX 77005}
\affiliation{Center for Theoretical Biological Physics, Rice University, Houston, TX 77030}
\affiliation{Departments of Chemistry and Physics \& Astronomy, Rice University, Houston, TX 77005}



\begin{abstract}

Tissues commonly consist of cells embedded within a fibrous biopolymer network. Whereas cell-free reconstituted biopolymer networks typically soften under applied uniaxial compression, various tissues, including liver, brain, and fat, have been observed to instead stiffen when compressed. The mechanism for this compression stiffening effect is not yet clear. Here, we demonstrate that when a material composed of stiff inclusions embedded in a fibrous network is compressed, heterogeneous rearrangement of the inclusions can induce tension within the interstitial network, leading to a macroscopic crossover from an initial bending-dominated softening regime to a stretching-dominated stiffening regime, which occurs before and independently of jamming of the inclusions. Using a coarse-grained particle-network model, we first establish a phase diagram for compression-driven, stretching-dominated stress propagation and jamming in uniaxially compressed 2- and 3-dimensional systems. Then, we demonstrate that a more detailed computational model of stiff inclusions in a subisostatic semiflexible fiber network exhibits quantitative agreement with the predictions of our coarse-grained model as well as qualitative agreement with experiments.

\end{abstract}


\maketitle
\section*{ INTRODUCTION}
Semiflexible biopolymer and fiber networks are well known for their unusual tendency to stiffen dramatically under applied shear or extensional strain  \cite{Fung1967, Gardel2004a, Storm2005, Tharmann2007, Kabla2007b, Picu2011, Broedersz2014RMP, Vahabi2016, VanOosten2016,  Ban2019} and soften under compression  \cite{Vahabi2016, VanOosten2016}. Many biological tissues, however, stiffen under applied compression \cite{Pogoda2014, Perepelyuk2016, VanOosten2019}, despite the fact that their structural backbone, the extracellular matrix, consists of an otherwise compression softening fiber network. Stiffening of tissues in response to uniaxial compression is ubiquitous in animals large enough to be subjected to gravitational stresses or other large forces. This behavior allows tissues to remain soft to small deformations needed for mechanosensing, while protecting them from damage induced by large compressive strains.  In addition, there is increasing evidence that cells sense and respond to compression-driven changes in tissue stiffness \cite{Pogoda2014}. This can have important consequences in, for example, brain tissue, which stiffens in response to increased blood pressure \cite{Arani2018} or the pressure gradient generated by a growing tumor \cite{Seano2019, Janmey2019}.

Whereas compression stiffening in tissues can be interpreted as a consequence of incompressibility of either the inclusions (cells) \cite{VanOosten2019} or the entire sample due to poroelastic effects \cite{Engstrom2019, Vahabi2016}, this behavior has also been demonstrated in biopolymer networks containing stiff (i.e. non-deforming) colloidal particles \cite{VanOosten2019}, for which the cause of compression stiffening is less clear. Developing a better understanding of the origin of this behavior, and in particular its dependence on the properties of both the inclusions and interstitial network,  may improve our knowledge about the nonlinear mechanics of tissues and support efforts to design functional biomimetic materials.

\begin{figure*}[!htb]
\centering
\includegraphics[width=1\textwidth]{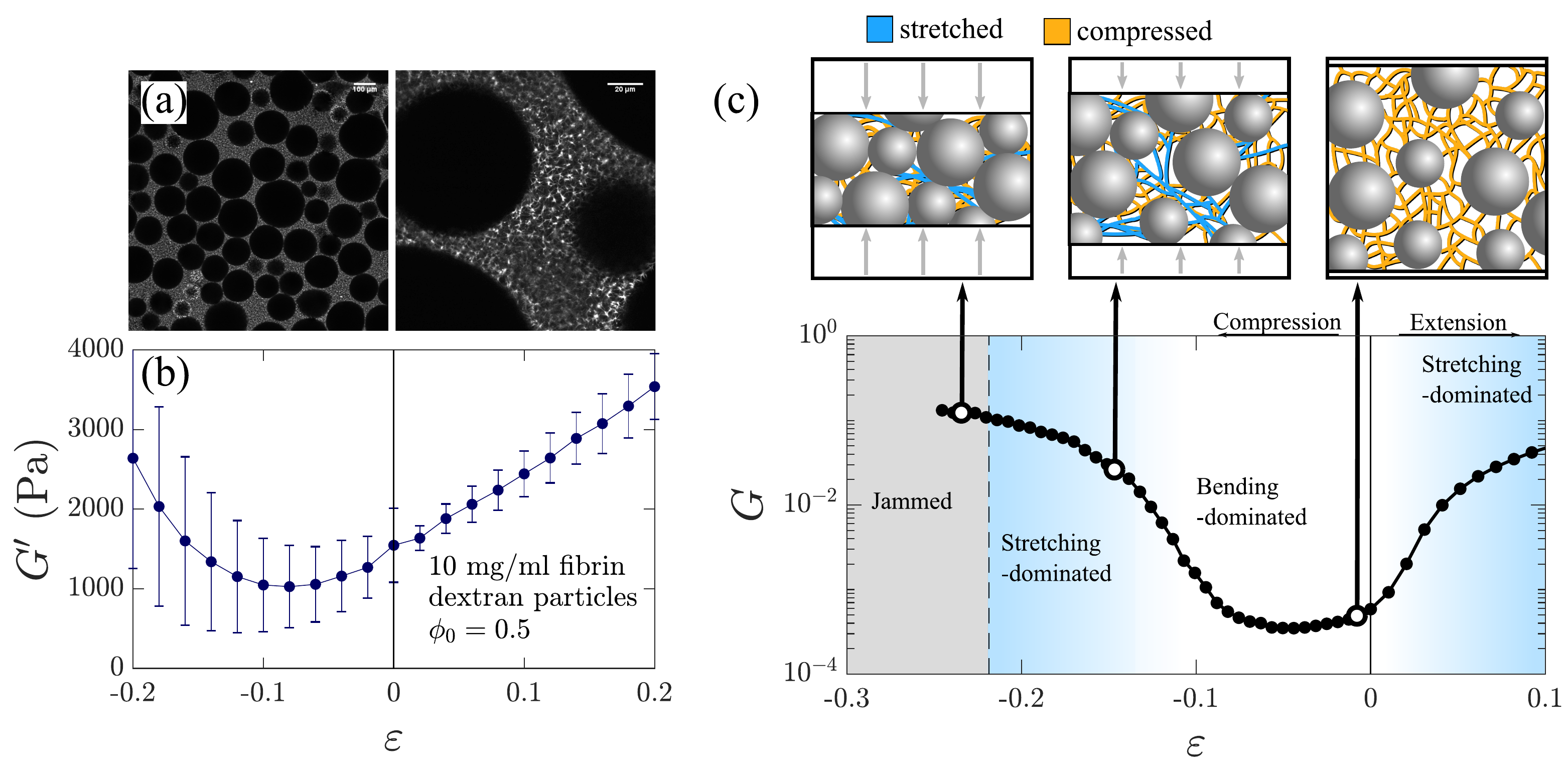}
\caption{\label{Figure1}  (a) Confocal images of a gel of 10 mg/ml fibrin containing stiff spherical dextran particles at a volume fraction of $\phi_0 = 0.5$. (b) Reproduced from Ref. \cite{VanOosten2019}: Shear storage modulus $G$ for the above system as a function of applied uniaxial strain. Under compression, the material initially softens before transitioning to a stiffening regime.  (c) Top right: Compression of a biopolymer network containing stiff inclusions initially leads to roughly uniform compression of the interstitial network (compressed fibers are colored orange), which leads to bending of the network fibers and softening of the macroscopic sample. Top middle: Further increasing compression leads to rearrangement of the stiff inclusions, which drives stretching of fibers in network regions between particles that move farther apart (stretched fibers are colored blue), leading to a macroscopic crossover from bending-dominated to stretching-dominated mechanics. Top left: With sufficiently large compression, the inclusions become jammed. Bottom: Schematic plot of the shear modulus $G$ as a function of applied uniaxial strain $\varepsilon$ for a simulated system of stiff particles embedded within a strain-stiffening network. The model is discussed in detail in Section III and this data appears in Fig. \ref{Figure4}.
}
\end{figure*}

In recent work, Van Oosten and coworkers measured the shear storage modulus, as a function of applied uniaxial strain, of a reconstituted fibrin network containing embedded stiff dextran particles  \cite{VanOosten2019}.  In Fig. \ref{Figure1}a, we reproduce their experimental data for samples with an initial inclusion volume fraction of $\phi_0 = 0.5$. Under increasing compression, this material exhibits an initial softening and subsequent stiffening regime. Notably, this unusual compression stiffening effect occurs while the volume fraction of the inclusions remains below the expected jamming threshold (see SI Appendix I). In contrast to these experiments with strain-stiffening biopolymer networks, the authors observed no compression stiffening effect below jamming in a system containing the same particles embedded within a linear elastic  (i.e. non-strain-stiffening) polyacrylamide gel. Thus, this unusual effect in the fibrin experiments appears to originate from some cooperative interaction between the mutual steric repulsion of the particles and strain stiffening properties of the network. In Fig. \ref{Figure1}c, we sketch a hypothetical mechanism for this behavior, along with a schematic plot of the shear modulus as a function of applied uniaxial strain. In a biopolymer network containing stiff inclusions, we expect that a small amount of applied macroscopic compression will result in homogeneous compression throughout the interstitial network, causing initial macroscopic softening akin to what is typically observed in compressed inclusion-free biopolymer networks \cite{VanOosten2016, Vahabi2016}. Inevitably, sufficient macroscopic compression of the sample induces contact and rearrangement of the sterically repulsive inclusions \cite{Shen2012}, driving local shear and extensional strain between neighboring inclusions. Provided that the critical extensional strain for stiffening of the interstitial network is sufficiently low that the magnitude of induced particle rearrangement induces local stiffening, this could lead to macroscopic stiff (tension-dominated) stress propagation before the inclusions become jammed.

In this work, we describe a new mechanism for compression stiffening in fibrous networks containing inclusions, which we show is related to nonaffine, cooperative particle rearrangement \cite{Lois2008, Shen2012} that occurs in random particle dispersions as macroscopic compression increases the particle volume fraction.  This rearrangement induces tension within the interstitial network, which in turn causes macroscopic stiffening. To explore the counterintuitive notion of compression-driven tension, we first consider the mechanics of a loosely distributed assembly of stiff, repulsive particles, in which neighboring particles are connected by soft springs that are \textit{rope-like}, meaning that they provide zero mechanical response to compression but behave as harmonic springs when stretched beyond a predefined slack extension. We refer to this as the \textit{rope model}, and treat it as a coarse-grained approximation of the zero-bending rigidity limit of a fibrous network containing inclusions.  Across a wide range of initial particle volume fractions, we find that applying sufficient uniaxial compression to this system induces a state of macroscopic stress propagation prior to jamming, in which stretching of the soft springs constitutes the dominant stress contribution. This is distinct from the jamming transition, which occurs at a well-defined particle volume fraction for a given shape and size distribution \cite{VanHecke2010,Koeze2016} and is dominated by compressive stress propagation \cite{cates1998jamming}. We find that stretching-dominated stress propagation appears to be related to contact percolation of the particles, which prior work has shown corresponds to the onset of increasing nonaffinity in the particle displacements in a macroscopically compressed particulate assembly \cite{Lois2008, Shen2012}. We generate phase diagrams for stretching-dominated stress propagation and jamming in 2D and 3D systems, as a function of both the particle volume fraction and the level of applied extension required for each ropelike spring to bear tension.

We then perform simulations of discrete disordered fiber networks, which have in prior work been shown to reproduce the nonlinear mechanical behavior of reconstituted biopolymer networks \cite{VanOosten2016, Vahabi2016, Licup2015}. In the absence of inclusions, these remain soft (mechanically bending-dominated) under applied compression and stiffen dramatically (becoming mechanically dominated by axial stretching of the network fibers) only when stretched beyond a critical extensional strain. We modify these model networks by embedding stiff, sterically repulsive particles that are rigidly connected to the surrounding network bonds. Similar simulations have been performed in recent work by Islam and coworkers \cite{Islam2019}, who showed that introducing rigid particles increases the linear modulus and reduces the extensional critical strain of strain stiffening networks. However, their work did not consider compression-driven phenomena. Here, we simulate the rheology of such networks under applied uniaxial strain and show that, with a sufficiently large volume fraction of embedded stiff inclusions, these exhibit significant compression stiffening, qualitatively reproducing the rheology of the experimental system. We show that this stiffening coincides with increasing nonaffine (heterogeneous) rearrangement of the inclusion positions. Further, we demonstrate that the volume-fraction dependence of this compression stiffening behavior is quantitatively captured by the predictions of the rope model.

\section*{RESULTS AND DISCUSSION}
\subsection*{Physical mechanism of compression-driven tension}

Biopolymer networks are unique in that they exhibit relatively weak, bending-dominated compressive response but stiffer, stretching-dominated tensile response above a critical applied strain. We hypothesize that the compression stiffening effect observed in particle-network composites is the result of tension within the the interstitial, strain-stiffening network caused by rearrangement of the sterically repulsive particles as the macroscopic sample is uniaxially compressed. This particle rearrangement is driven by the inability of the non-deforming, spherical particles to accomodate a homogeneous deformation field under uniaxial compression due to their mutual steric repulsion. In this section, we consider a coarse-grained model consisting of a random arrangement of stiff repulsive particles, in which  neighboring particles are connected by soft, rope-like springs that are harmonic under applied extension but have no resistance to compression. In this zero-bending limit, we demonstrate that compression-driven particle rearrangement can induce stretching-dominated, sample-spanning stress propagation, at a volume-fraction-dependent critical compression prior to jamming.

\begin{figure*}[!htb]
\centering
\includegraphics[width=1\textwidth]{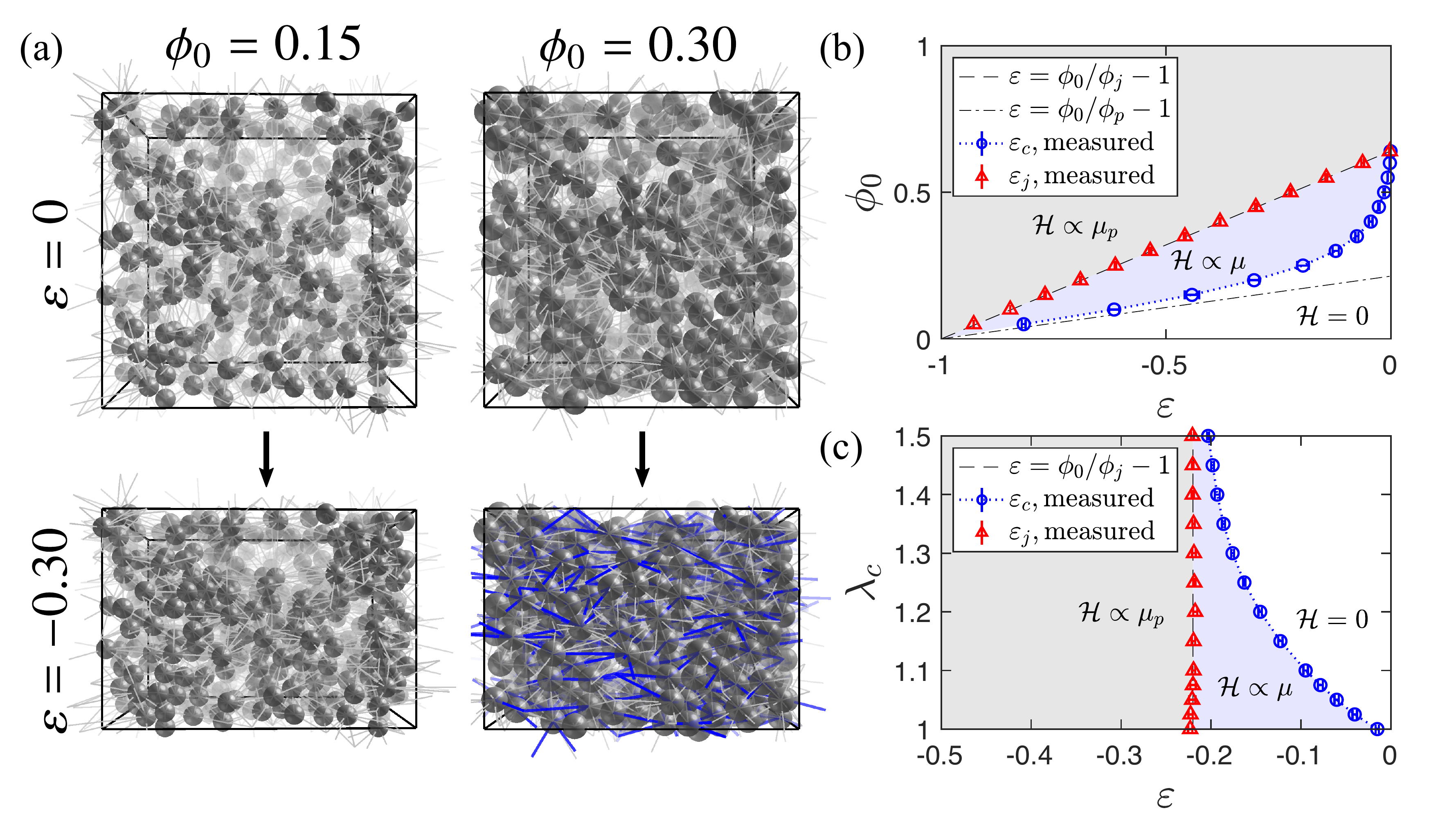}
\caption{\label{Figure2} (a) We apply uniaxial compression to periodic systems comprised of $N = 1000$ randomly placed, repulsive spherical particles with initial volume fractions $\phi_0 = 0.15$ (left) and $\phi_0 = 0.3$ (right), in which neighboring particles are connected according to the Delaunay triangulation of the particle centers. The springs connecting pairs of particles are rope-like, meaning that they only produce finite tension when stretched beyond a slack extension $\lambda_c$. Applying a uniaxial compression of $\varepsilon = -0.3$ to a 3D system with initial volume fraction $\phi_0=0.3$ results in sufficient rearrangement of the sterically repulsive particles to induce sample-spanning, tension-dominated stress propagation (blue springs are stretched), whereas the same compression is insufficient to stress propagation for $\phi_0 = 0.15$. (b) Mechanical phase diagram for compressed systems of $N = 1000$ repulsive spheres of modulus $\mu_p = 1$, in which nearest neighbors (by Delaunay triangulation) are connected by rope-like springs of modulus $\mu = 10^{-5}$, as a function of uniaxial strain $\varepsilon$ and initial volume fraction $\phi_0$. Here, the rope-like springs have critical extension $\lambda_c = 1$. The blue circles correspond to the critical strain for the onset of tension propagation, $\varepsilon_c$, and the red triangles correspond to the onset of jamming, $\varepsilon_j$. The dashed black line corresponds to the predicted applied strain required for jamming of a system with initial volume fraction $\phi_0$, $\varepsilon_j = \phi_0/\phi_j - 1$, in which $\phi_j = 0.64$. The white region corresponds to the floppy regime ($\mathcal{H}_{\mathrm{total}}=0$, blue corresponds to the stretching-dominated regime ($\mathcal{H}_{\mathrm{total}}\propto\mu$), and gray corresponds to the jammed regime ($\mathcal{H}_{\mathrm{total}}\propto\mu_p$). (b) Mechanical phase diagram for volume fraction $\phi_0 = 0.5$ as a function of applied uniaxial strain $\varepsilon$ and slack extension of rope-like springs, $\lambda_c$. Error bars in both panels correspond to $\pm1$ standard deviation. 
} 
\end{figure*}

We consider random arrangements of $N$ radially monodisperse spheres (bidisperse disks in 2D, with a ratio of radii of 1.4 to avoid crystallization \cite{Koeze2016}, in which the two particle size subpopulations are equal in number), in a periodic box of volume $L^d$, in which $L$ is chosen so that the initial particle volume fraction (area fraction in 2D) is $\phi_0$. Further details on sample generation are provided in Methods. We use $N = 1000$ in 3D and $N = 900$ in 2D.  Neighboring particles, as identified by the Delauney triangulation of the particle centers, are connected by rope-like springs.  The initial lengths of the rope-like springs are set to be equal to the initial distance between each pair of neighboring particles, such that an increase in the distance between two nearest neighbors corresponds to extension of the pair’s connecting spring. Because we intend to treat each spring as a coarse-grained approximation of a strain-stiffening network region between each pair of inclusions,  we define a “slack extension” $\lambda_c$ above which each rope-like spring transmits tension. The energy for the rope-like springs is written as follows

\begin{equation}
\mathcal{H}_{\mathrm{rope}} = \frac{\mu}{2}\sum_{ ij }{\frac{\left(\ell_{ij}-\lambda_c\ell_{ij,0}\right)^2}{\lambda_c\ell_{ij,0}}\Theta\left(\ell_{ij}-\lambda_c\ell_{ij,0}\right)}
\end{equation}
in which $\mu$ is the spring constant, $\Theta$ is the Heaviside step function, $\ell_{ij}$ is the distance between the centers of particles $i$ and $j$, and the sum is taken over all springs between neighboring particles. The repulsive energy between overlapping particles is written as

\begin{equation}
\mathcal{H}_{\mathrm{repulsion}} = \frac{\mu_p}{2}\sum_{mn}{\left(1-\frac{\ell_{mn}}{r_{mn}}\right)^2\Theta\left(1-\frac{\ell_{mn}}{r_{mn}}\right)}
\end{equation}
in which $\mu_p$ is the one-sided repulsive spring constant and the sum is taken over all pairs of particles $m$ and $n$. As we are interested in the limit in which the particles are much stiffer than the springs, we set $\mu=10^{-5}$ and $\mu_p=1$. Again, this system can be thought of as a coarse-grained description of a biopolymer network containing embedded particles, which ignores the relatively weak,  bending-dominated linear and compressive mechanical responses of the network and instead considers both repulsion between overlapping stiff particles and the stretching-dominated mechanics of network regions driven above the critical strain.

To this system, we apply quasistatic uniaxial strain $\varepsilon$ in small steps ($|d\varepsilon| < 0.01$) using generalized Lees-Edwards periodic boundary conditions \cite{Lees1972}. At each step, we minimize the total energy $\mathcal{H}_{\mathrm{total}} = \mathcal{H}_{\mathrm{rope}} + \mathcal{H}_{\mathrm{repulsion}}$ using the L-BFGS method \cite{Nocedal2006}. In Fig. \ref{Figure2}a, we plot sample configurations for a 3D system with $N=1000$ particles at initial volume fractions of $\phi_0 = 0.15$ and $\phi_0 = 0.30$, under varying levels of compression, in which the slack extension for the rope-like springs is set to $\lambda_c = 1$, such that any stretching results in a nonzero energy contribution. Under an applied compression of $\varepsilon = -0.3$, no stretching is induced in the system with $\phi_0 = 0.15$, so it remains mechanically floppy, with $\mathcal{H}_{\mathrm{total}} = 0$. When the same amount of compression is applied to the sample with $\phi_0 = 0.3$, however, compression-driven rearrangement of the sterically repulsive particles induces sample-spanning stress propagation in which the mechanics are dominated by stretching, leading to $\mathcal{H}_{\mathrm{total}}\propto\mu$. The stretching of a significant fraction of the bonds is evident in the bottom right panel of Fig. \ref{Figure2}a, in which stretched bonds are colored blue. Importantly, the stretching-dominated stress propagation shown in Fig. \ref{Figure2}a occurs at a lower level of applied compression than that required for jamming; thus in the absence of the springs, the system would be a floppy, unjammed particle assembly. Sufficiently increasing the compression applied to a given system eventually leads to jamming, i.e. stress propagation dominated by repulsive forces between particles, such that $\mathcal{H}_{\mathrm{total}}\propto\mu_p$, at $\varepsilon_j = \phi_0/\phi_j - 1$. Here, $\phi_j$ is the jamming volume fraction of packings of spheres 3D ($\phi_{j}\approx0.64$) and radially bidisperse disks 2D ($\phi_{j}\approx0.84$). Note that in our calculation of $\varepsilon_j$, we assume that the particle volume fraction at the onset of jamming is the same for packings under uniaxial compression as for packings under bulk compression, under which $\phi_j$ is commonly measured. While this assumption may not be strictly true, we find that it works well for our simulations.

To explore the full volume-fraction dependence of this effect, in Fig. \ref{Figure2}b, we plot a mechanical phase diagram for compressed 3D systems with slack extension $\lambda_c = 1$ and varying initial volume fraction $\phi_0$. We show the same phase diagram for 2D systems in Fig. S1 in SI Appendix II.  We identify the critical compressive strain corresponding to stretching-dominated stress propagation, $\varepsilon_c$, as the first applied strain in which the system's energy becomes finite (we choose a threshold of $\mathcal{H}_{\mathrm{total}}/V=10^{-11}$). That this initial stress propagation is stretching-dominated is evident from the fraction of the total energy deriving from stretching, $\mathcal{H}_{\mathrm{rope}}/\mathcal{H}_{\mathrm{total}}$, which is approximately 1 at the onset of finite $\mathcal{H}_{\mathrm{total}}$. We identify the critical compressive strain for jamming, $\varepsilon_j$, as the strain at which the fractional contribution of repulsion to the total energy, $\mathcal{H}_{\mathrm{repulsion}}/\mathcal{H}_{\mathrm{total}}$, exceeds 0.5. We find that the measured critical strains for jamming obey the predicted dependence on the initial volume fraction, $\varepsilon_j = \phi_0/\phi_j - 1$, in both 2D and 3D, and we observe that tension propagation occurs prior to jamming over a wide range of initial volume fractions.  Near $\phi_0 = 0$, we find that the critical strain for tension propagation with $\lambda_c = 1$ seems to approximately match the expected applied strain for contact percolation of particles with short-range attractive interactions, $\varepsilon_p = \phi_0/\phi_p - 1$, where $\phi_p\approx 0.214$ in 3D and $\phi_p \approx 0.558$ in 2D \cite{Shen2012}. Note that this calculation of $\varepsilon_p$ has the same caveat as our prediction of $\varepsilon_j$, in that the referenced values of $\phi_p$ were measured under bulk compression. Perhaps surprisingly, we observe that stress propagation occurs slightly before the contact percolation point for an intermediate range of volume fractions in 2D. We discuss this in SI Appendix II (see Fig. S2).

Intuitively, for a fixed volume fraction, increasing the slack extension $\lambda_c$ of the ropelike springs should increase the amount of compression-driven rearrangement required for tension propagation. Consequently, $\varepsilon_c$ should become more negative with increasing $\lambda_c$. In contrast, the required compression for jamming, $\varepsilon_j$, strictly depends on $\phi_0$ and is not expected to show any dependence on $\lambda_c$. In Fig. \ref{Figure2}c, we plot $\varepsilon_c$ and $\varepsilon_j$ for 3D systems with $\phi_0 = 0.5$ and varying $\lambda_c$, and we plot the complementary data for 2D systems in Fig. S1 in SI Appendix II. We plot $\varepsilon_c(\lambda_c)$ for a several volume fractions in both 2D and 3D in Fig. S3.

This compression-driven, tension-dominated stress propagation is a  geometric effect, caused by the inevitably heterogeneous rearrangement of a compressed assembly of sterically repulsive particles as the particle volume fraction increases toward jamming. Thus, the shape of the phase boundaries in Fig. \ref{Figure2}b-c should not depend on the precise elastic properties of the rope-like springs and particles provided that $\mu_p \gg \mu$. If, in fibrous networks containing embedded stiff particles, the dominant modes of stretching under macroscopic compression correspond to tension between nearest neighbor particles, then the phase boundaries in Fig. \ref{Figure2}b-c may be useful for predicting the onset of compression stiffening in strain-stiffening networks containing inclusions, such as the example depicted in Fig. \ref{Figure1}a, provided that the inclusion volume fraction and extensional critical strain of the underlying fiber network are known. In the following section, we test these ideas using simulations of disordered elastic networks containing stiff inclusions.

\subsection*{Model of a strain-stiffening network containing stiff inclusions}

This rope model has suggested a new mechanism for compression-driven, stretching-dominated stress propagation in strain-stiffening materials containing sterically repulsive particles. To establish the validity of this mechanism, we now perform simulations using a more established fiber network model containing rigid particles, for which we can measure the influence of  applied uniaxial strain (compression or extension) on the linear shear modulus. 

Prior work has demonstrated that the mechanics of semiflexible polymer networks are strongly influenced by the connectivity $z$, defined as the average number of bonds connected to a network node \cite{Broedersz2011, Broedersz2014RMP}. A network of initially unstressed and athermal Hookean springs with 1D modulus $\mu$ has a finite shear modulus $G\propto\mu$ only if the average connectivity $z$ is equal to or greater than an \emph{isostatic} threshold $z_c = 2d$, identified by Maxwell, where $d$ is the dimensionality \cite{Maxwell1864}. The addition of soft bending interactions with modulus $\kappa$ results in a bending-dominated regime with  $G \propto \kappa$ for $z < z_c$, with a crossover to a stretching-dominated regime with $G\propto\mu$ for $z > z_c$ \cite{Head2003,Wilhelm2003,Onck2005,Broedersz2011}.
For extracellular matrices of collagen or fibrin, the average connectivity $z\lesssim4$ is well below the 3D isostatic threshold of 6 \cite{Vader2009,Jansen2018}.
Thus, if such \emph{subisostatic} networks are athermal, it is the former bend-dominated regime that is expected to describe the linear elastic modulus.
In this linear, bending-dominated mechanical regime, the introduction of tensile prestress (e.g., by molecular motors \cite{Broedersz2011a,Sheinman2012a} or applied extension \cite{Vahabi2016, VanOosten2016, Cui2019}) drives an increase in the shear modulus. In fact, sufficiently large applied shear or extensional strain \cite{Sharma2016a, Sheinman2012} can induce a crossover to a stiff, stretching-dominated regime \cite{Storm2005, Onck2005, Sharma2016a, Vahabi2016, VanOosten2016}, with the magnitude of required strain decreasing to zero as $z\to z_c$ \cite{Wyart2008}. In contrast, under applied compression, networks typically soften relative to the unstrained state \cite{Cui2019}, remaining (in the case of biopolymer networks) within the bending-dominated regime \cite{Vahabi2016, VanOosten2016, VanOosten2019}.   In this section, we demonstrate that embedding repulsive particles within such networks leads to compression stiffening, at a level of compression that is controlled by a combination of the network critical strain and the particle volume fraction. We find that the phase diagram for the rope model discussed in the previous section quantitatively captures the volume-fraction dependence of the compression stiffening effect in this more realistic model. 

We generate periodic, subisostatic fiber networks of average connectivity $z$ derived from dense 3D sphere packings, as described in Methods and SI Appendix III. Then, we randomly place non-intersecting spherical inclusions of radius $r$ within the simulation box until the desired inclusion volume fraction $\phi_0$ is reached. Any network bond that intersects with the boundary of an inclusion is connected to the inclusion surface, at the intersection point, by a freely rotating joint, and all remaining bond segments with the inclusion boundary are removed. Each inclusion transforms as a rigid object with (in 3D) $3$ translational and $3$ rotational degrees of freedom. Example images of a network containing inclusions, prior to applied deformation, are provided in Fig. S5 in SI Appendix III.

 For a given configuration, the total energy $\mathcal{H} $ of the system is computed as
\begin{equation}
\mathcal{H} = \mathcal{H}_S  + \mathcal{H}_B + \mathcal{H}_R,
\end{equation}
 in which $\mathcal{H}_S$, $\mathcal{H}_B$, and $\mathcal{H}_R$ represent stretching, bending, and repulsive contributions, respectively. We treat individual segments as harmonic springs of modulus $\mu$, compute harmonic bending interactions of modulus $\kappa$ between nearest-neighbor segments, and account for one-sided harmonic repulsive interactions of modulus $\mu_R$ between pairs of inclusions and between inclusions and network nodes. Unless stated otherwise, we set $\mu  = 1$ and $\kappa = 10^{-4}$ so that the linear elasticity of the interstitial, subisostatic network is bending-dominated. Prior work has shown that networks with $\kappa$ around this magnitude reasonably capture the mechanical behavior of reconstituted collagen and fibrin networks \cite{Licup2015, VanOosten2016, Vahabi2016, Jansen2018}. We set $\mu_R = 100$ so that the repulsive interactions are significantly stiffer than both the bending and stretching interactions.  Further details are provided in Methods. Since we focus on the regime below jamming ($|\varepsilon|<|\varepsilon_j|$) throughout this work, our results should be qualitatively consistent with $\mu_R \to \infty$ provided that $\mu_R\gg \{\mu,\kappa\}$.   Using the procedure described in Methods, we measure the linear shear modulus $G$ as a function of uniaxial strain $\varepsilon$ for compressive and extensional strains over a range of inclusion volume fractions.
	
\begin{figure}[!htb]
\centering
\includegraphics[width=0.8\columnwidth]{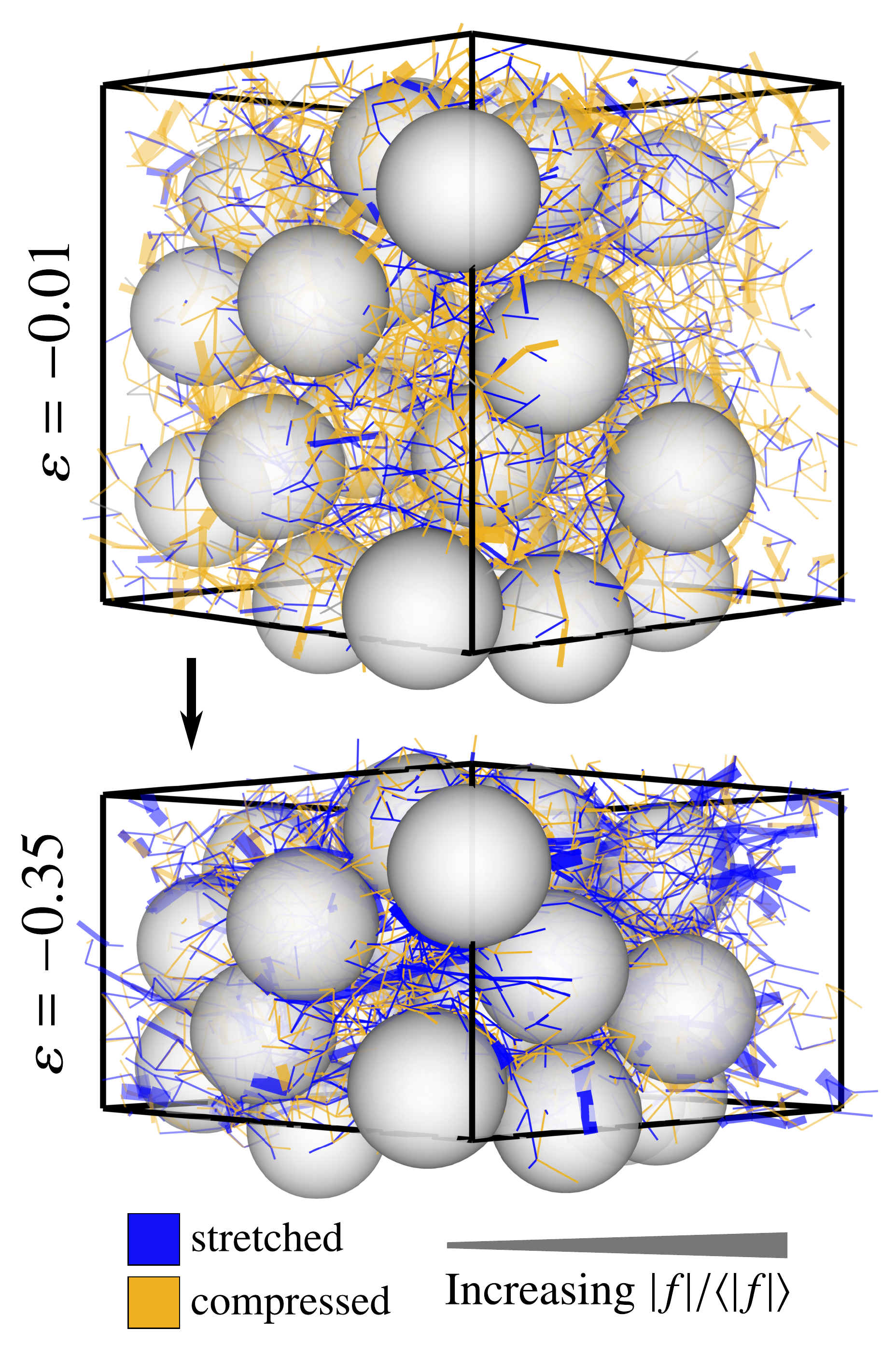}
\caption{\label{Figure3} Images of a periodic packing-derived network unit cell with $L = 15$, $z = 4$, $\kappa = 10^{-4}$, and spherical inclusions with radius $r = 2$ and volume fraction $\phi_0 = 0.3$ under varying levels of compression. Compressed bonds are colored orange and stretched bonds are  blue. Bond thickness is proportional to the magnitude of the tensile/compressive force $|f|$ on the bond normalized by the average force magnitude $\langle| f |\rangle $, with thresholds at $|f|/\langle |f| \rangle = [1,8]$. The dimensions of the outer box represents the periodic Lees-Edwards boundary conditions.
}
\end{figure}

\begin{figure*}[!htb]
\centering
\includegraphics[width=1\textwidth]{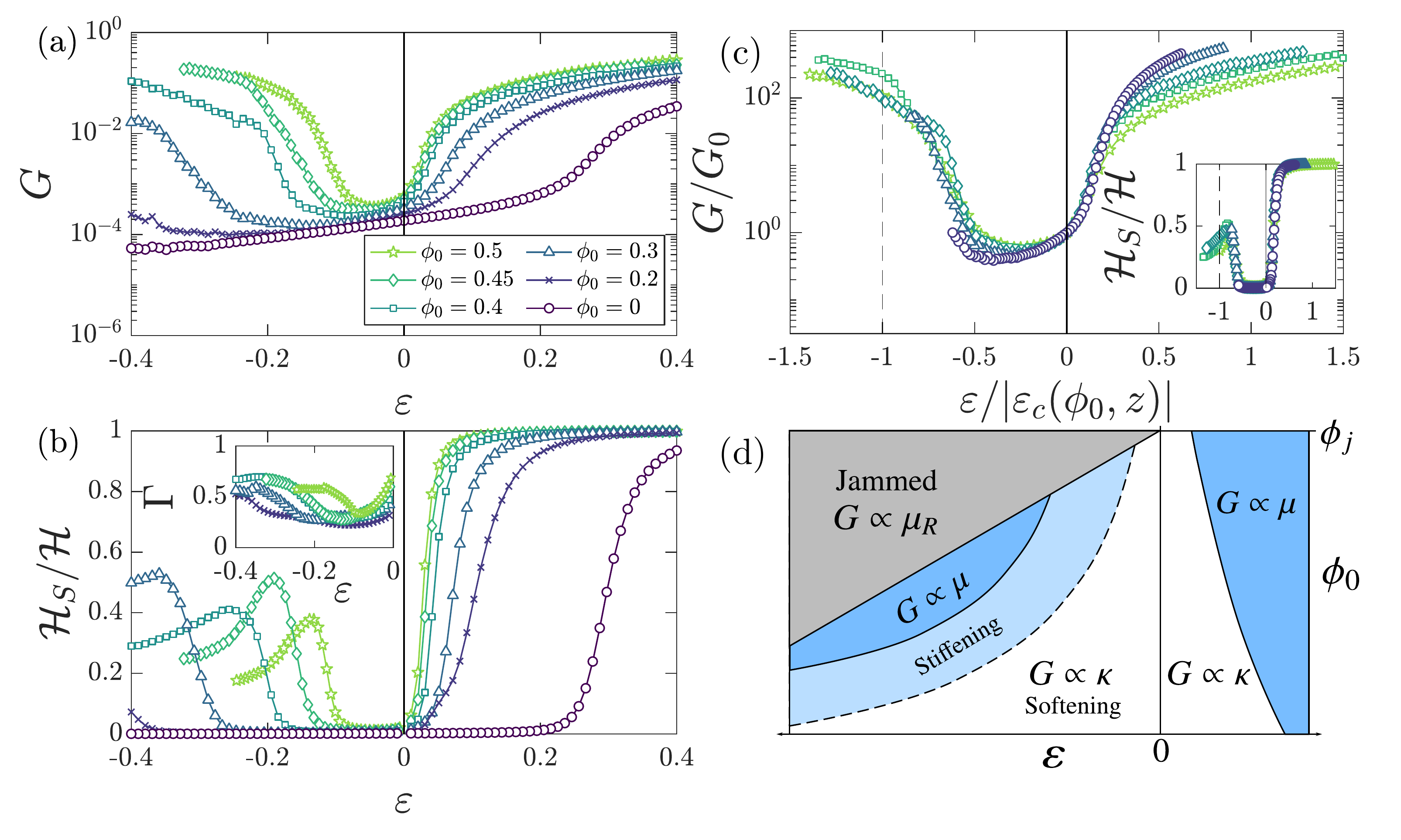}
\caption{\label{Figure4} (a) Shear modulus $G$ as a function of applied uniaxial strain $\varepsilon$ for 3D packing-derived networks with $z = 4$, $\kappa = 10^{-4}$, and varying initial inclusion volume fraction $\phi_0$. Networks with large enough $\phi_0$ undergo a compression-driven crossover from a bending-dominated softening regime to a stiffening regime and eventually become stretching-dominated, as reflected by the (b) stretching energy fraction $\mathcal{H}_S/\mathcal{H}$. With increasing $\phi_0$, the crossover point occurs at lower levels of applied compression. Inset: Nonaffinity $\Gamma$ of the inclusion deformation as a function of applied compression. (c) Data from (a) normalized by the zero-strain shear modulus $G_0 = G(\varepsilon=0$) and plotted as a function of the uniaxial strain $\varepsilon$ normalized by the predicted critical strain for compression-driven tension propagation, $\varepsilon_c(\phi_0,z)$, where $\varepsilon_{c,ext}=0.3$ is the critical \textit{extensional} strain for empty networks with $z = 4$, and $\varepsilon_c(\phi_0,\lambda_c = 1 + \varepsilon_{c,ext})$ is determined from the 3D rope model, as described in SI Appendix II. (d) Schematic phase diagram for the shear modulus $G$ as a function of applied uniaxial strain $\varepsilon$ and initial inclusion volume fraction $\phi_0$.} 
\end{figure*}

	First, we consider networks with $z = 4$ which, without inclusions, soften under compression but stiffen at a critical extensional strain of $\varepsilon_{c,ext}\approx0.3$, identified as the inflection point of the $G$ vs. $\varepsilon$ curve for $\phi_0 = 0$ in Fig. \ref{Figure4}a. When inclusions are present with a sufficient $\phi_0$, these stiffen under both applied extension and compression. In Fig. \ref{Figure4}a, we plot the shear modulus for the same networks with varying initial inclusion volume fraction $\phi_0$ . We find that networks containing sufficiently large $\phi_0$ undergo a compression softening regime at low levels of compression followed by stiffening at higher levels of compression, similar to the behavior observed in the experiments of Ref. \cite{VanOosten2019} shown in Fig. \ref{Figure1}a.  In Fig. \ref{Figure3}, we provide images of a simulation with $\phi_0 = 0.3$ and $z = 4$ under varying levels of applied compression. Whereas most bonds are compressed (orange) at the relatively low applied macroscopic compression of $\varepsilon = -0.01$,  at the more substantial compression of $\varepsilon = -0.35$ we observe significant stretching (blue) of network regions between neighboring inclusions. As in the rope model, this stretching is driven by rearrangement of the sterically repulsive inclusions as the system approaches jamming. We find that increasing $\phi_0$ leads to a decrease in the magnitude of applied compression corresponding to the minimum in $G$, beyond which the networks stiffen with increasing compression. For sufficiently large $\phi_0$ and sufficient applied compression, these enter a stretching-dominated stiffening regime with $G\propto\mu$ at a critical compressive strain that decreases with increasing $\phi_0$, in qualitative agreement with the phase diagram in Fig. \ref{Figure2}d.  In Fig. \ref{Figure4}b, we plot the stretching energy fraction $\mathcal{H}_S/\mathcal{H}$ as a function of strain for the same networks, demonstrating that compression stiffening coincides with a crossover from a bending-dominated regime to a stretching-dominated regime. To emphasize this point, we repeat these measurements for networks with varying the bending modulus $\kappa$ and fixed $\phi_0 = 0.4 $ and $z = 4$ (see Fig. S6 in SI Appendix III). These show a clear shift from a softening regime in which $G\propto\kappa$ at small strains, to a crossover stiffening regime at intermediate strains, to a stretch-dominated stiffening regime with $G\propto\mu$  at larger strains. In Fig. \ref{Figure4}d, we draw a schematic phase diagram for the shear modulus of a strain-stiffening fibrous network containing rigid inclusions as a function of inclusion volume fraction and applied uniaxial strain. In Fig. S8 in SI Appendix III, we replot the data from Fig. \ref{Figure4}, colored by the magnitude of $G$, over a plot of $\phi_0$ vs. $\varepsilon$, revealing the regimes sketched in Fig. \ref{Figure4}d.

To explore the influence of strain heterogeneity on the compression stiffening effect, we measure the nonaffinity $\Gamma(\varepsilon)$ of the inclusion deformation field as follows
\begin{equation}
\Gamma = \frac{1}{r^2\varepsilon^2}\big\langle \left|\vec{\bm{u}}_i (\varepsilon)-\vec{\bm{u}}_{i,{\mathrm{aff}}}(\varepsilon)\right|^2\big\rangle_i
\end{equation}
in which the average is taken over all inclusions, $r$ is the inclusion radius, $\vec{\bm{u}}_i(\varepsilon)$ is the actual position of inclusion $i$ after relaxation under applied strain $\varepsilon$, and $\vec{\bm{u}}_{i,{\mathrm{aff}}}(\varepsilon)$ is the position of inclusion $i$ under affine (homogenous) deformation of the initial network configuration by strain $\varepsilon$. Note that since our simulations are periodic, we shift $\vec{\bm{u}}_i$ and $\vec{\bm{u}}_{i,{\mathrm{aff}}}$ for the purposes of calculating $\Gamma$ so that the average of each corresponds to the origin. Under an applied compression of $\varepsilon$, particles are displaced by an average distance $d_\Gamma = r|\varepsilon|\sqrt{\Gamma}$ from their expected locations under affine deformation. As $\Gamma$ is proportional to the typical squared distance of each inclusion from its expected position under macroscopic homogenous compression, increasing rearrangement of the inclusions will be indicated by increasing $\Gamma$.  In the inset of Fig. \ref{Figure4}b, we plot $\Gamma$ for systems with varying $\phi_0$. We find that $\Gamma$ decreases in all systems in the compression softening regime, but it increases throughout the compression stiffening regime, beginning at roughly the same strains in which the shear modulus $G$ begins to increase. As $\Gamma$ can in principle be measured in experiments via particle-tracking, this quantity could be used to test whether increasing strain heterogeneity drives compression stiffening in experiments. We note that, given the cooperative nature of inclusion rearrangement, $\Gamma$ is expected to increase in magnitude with increasing system size (number of inclusions), which is relatively limited for our 3D simulations. In Fig. S10 in SI Appendix III, we show that in larger 2D simulations, $\Gamma$ grows more dramatically than in the smaller 3D systems.

In the previous section, for a random configuration of particles with initial volume fraction $\phi_0$, with neighboring particles connected by rope-like springs with slack extension $\lambda_c$, we determined the critical compressive strain $\varepsilon_c$ for stretching-dominated stress propagation. Since the rope-like springs act as a coarse-grained approximation of a strain-stiffening network with a known critical extensional strain $\varepsilon_{c,ext}(\phi_0=0,z)$, we should compare our simulations to the rope model with slack extension $\lambda_c = 1 + \varepsilon_{c,ext}$.
Using $\varepsilon_c(\phi_0,\lambda_c)$ determined for the rope model, as shown in Fig. S3 (see SI Appendix II), we can thus predict the critical compressive strain $\varepsilon_c(\phi_0,z)\equiv\varepsilon_c(\phi_0,\lambda_c=1+\varepsilon_{c,ext}(\phi_0=0,z))$ for stretching-dominated stress propagation in our simulations. In Fig. \ref{Figure4}c, we show that normalizing $G(\varepsilon)$ in Fig. \ref{Figure4}  by the zero-strain shear modulus $G_0 = G(\varepsilon = 0)$ and normalizing the applied uniaxial strain by the predicted critical compressive strain $\varepsilon_c(\phi_0,z)$ leads to reasonable collapse of the data from Fig. \ref{Figure4}a under compression. In Fig. \ref{Figure4}c (inset), we show  that the stretching energy fraction $\mathcal{H}_S/\mathcal{H}$ curves also collapse when plotted as a function of $\varepsilon/\varepsilon_c(\phi_0,z)$. 

Based on our observation that increasing the slack extension of the springs in the rope model leads to an increase in the magnitude of the critical compression required for stretching-dominated stress propagation, we anticipate that increasing $z$ should decrease the magnitude of compression required for stiffening in our simulated network-inclusion composites.  In inclusion-free, subisostatic networks, increasing $z$ results in a decrease in the critical applied extensional strain required for stiffening.  In simulations, we find that for systems with a fixed $\phi_0$, decreasing the extensional critical strain of the underlying network by increasing $z$ in networks results in a decrease in the magnitude of applied compression at which the system stiffens (see Fig. S7a in SI Appendix III) and becomes stretching-dominated (see Fig. S7b in SI Appendix III), in agreement with our expectations based on the rope model.

For the case of applied extension, we find that our results agree qualitatively with those of Islam and coworkers, who used a similar model of random 3D networks containing inclusions \cite{Islam2019}. Under increasing extension, we observe an initial bending-dominated stiffening regime, with $G\propto\kappa$, followed by a crossover to a much stiffer stretching-dominated regime, with $G\propto\mu$, in agreement with their results for bonded inclusions \cite{Islam2019}.  This transition occurs at a critical extensional strain that decreases with increasing $\phi_0$, in further agreement with their work. Likewise, we find that the volume-fraction-dependence of the critical extensional strain is sensitive to the nature of the connections between the inclusions and network. In Fig. S12 in SI Appendix III, we consider inclusions that are disconnected from the surrounding network, and in this case we find that increasing $\phi_0$ leads to an increase in the critical extensional strain (see Fig. S12), as observed in Ref. \cite{Islam2019}.  However, we find that these, too, stiffen under compression at a volume-fraction-dependent critical strain, in qualitative agreement with our other results.

In Fig. S10 in SI Appendix III, we plot analogous data for Fig. \ref{Figure4}a for 2D packing-derived networks with $z = 3$ and varying $\phi_0$. These exhibit compression stiffening behavior that qualitatively agrees with our 3D simulations. 

\section*{CONCLUDING REMARKS}

We have demonstrated that the compression-driven nonaffine rearrangement of stiff particles embedded in a network can generate tension-dominated stress propagation, coinciding with macroscopic compression stiffening, and we have shown how this effect is controlled by both the volume fraction of particles and the strain-stiffening properties of the interstitial network. Using simulations of disordered 3D elastic networks containing stiff inclusions, which qualitatively reproduce the compression stiffening behavior observed in experiments, we have provided evidence that, given both the volume fraction of inclusions and the critical extensional strain of the interstitial strain stiffening network, one can utilize the rope model to predict the critical compressive strain corresponding to stretching-dominated stress propagation. Our results suggest a strategy for rational design of nonlinear mechanics in engineered tissues or synthetic composite materials using controlled volume fractions of inclusions. 

Interestingly, a similar rearrangement-driven stiffening effect might occur below the jamming transition for networks containing deformable inclusions, such as cells, provided that these sufficiently resist changes in volume. If so, then the rearrangement-driven stretching effect described in this work may drive the compression stiffening behavior that has been widely observed in living tissues. To study a more direct model of tissues, one could replace the stiff inclusions in our model with deformable particles that resist area/volume change \cite{Delarue2016, Boromand2018} and/or actively exert forces on the network \cite{Liang2016, Zhang2018}. Contractile cells, for example, might suppress or entirely remove the initial compression softening effect by pre-stretching regions of the interstitial network. Further, large scale force generation induced by contractile cells \cite{Ronceray2016,Han2018} may enhance the rearrangement-driven stiffening described in our work. Indeed, the authors in Ref. \cite{VanOosten2019} showed that fibrin networks containing particles and cells stiffen more than networks containing particles alone. Prestress may also explain why an initial compression softening regime is not typically observed in tissue samples \cite{Pogoda2014, Perepelyuk2016, VanOosten2019}.

While our simulations assume that bonds are permanent, real biopolymers rupture under sufficiently large extension. In Fig. S11, we measure the maximum bond extension $\max(\ell_{ij}/\ell_{ij,0})$ in a 3D sample with $\phi_0 = 0.5$ and $z = 4$, and find that even at the point of jamming, the maximum stretch does not exceed $50\%$.  Since this work is motivated by experiments on fibrin, which can be stretched far beyond this value, we do not include effects of rupture. Nevertheless, this effect may be relevant for less extensible biopolymers, such as collagen. In addition, as we focus on the quasistatic limit in this work, we ignore any mechanical influence of the fluid (i.e. poroelastic effects, surface tension, etc.). These may become important in samples with large particle volume fractions. Future work will need to include such effects, particularly if the dynamic properties of the material are to be considered. 

Whereas we have considered only the case of zero lateral strain in this study, we note that different boundary conditions could certainly affect our results. For example, a uniaxially compressed sample with free lateral boundaries would not necessarily jam at $\varepsilon_j$, but rather the inclusions would instead rearrange by spreading outward to accomodate increasing compression. Nevertheless, this effect would lead to increasing shear/extension in the regions between inclusions, so we expect that this would simply increase the duration of the stretching-dominated stiffening regime and delay the onset of jamming. This is possibly relevant to experimental results in Ref. \cite{VanOosten2019} on fibrin networks with dextran inclusions at $\phi_0 = 0.6$. Although $\varepsilon_j \approx -0.06$ for $\phi_0 = 0.6$, these were shown to continue to compression stiffen at even larger compressive strains of up to $\varepsilon = -0.2$.

The rope model, given its simple nature, lends itself easily to further exploration.
One could, for example, replace the rope-like springs between neighbors with springs that follow a more complex force-extension curve, e.g. that of extensible thermal worm-like chains \cite{Storm2005}.

\section*{METHODS}
\subsection*{Generation of coarse-grained rope model}

As described in the main text, we consider $N$ radially monodisperse spheres in a periodic box of volume $L^d$, in which $L$ is chosen such that the initial particle volume fraction (area fraction in 2D) is $\phi_0$.  We use $N = 1000$ in 3D and $N = 900$ in 2D.  To generate the initial particle configuration, we first randomly choose $N$ initial locations as particle centers and increase the particle radii from 0 in small steps, allowing the system’s energy to relax at each step using the L-BFGS method \cite{Nocedal2006} to avoid particle overlap. Upon reaching the desired radii, we generate the Delaunay triangulation of the particle centers \cite{CGAL} to identify pairs of neighboring particles, which we subsequently connect with rope-like springs.  The initial lengths of the rope-like springs are set to be equal to the initial distance between each pair of neighboring particles.

\subsection*{Generation of subisostatic networks containing stiff inclusions}

We begin with a packing-derived network composed of $N$ nodes with average connectivity $z_0\approx8$ in a 3D periodic unit cell of volume $V = L^3$. We then randomly delete bonds until the desired average network connectivity $z$ is realized. We randomly place inclusions of radius $r = 2$ with total volume fraction $\phi_0$ within the periodic box, connecting these to the network at points of intersection by freely rotating joints. Further details are provided in SI Appendix III. Unless otherwise stated, we use $L=15$ and $N=15^3$.

The total energy $\mathcal{H}$ of the network is computed as
\begin{equation}
\mathcal{H} = \mathcal{H}_S + \mathcal{H}_B + \mathcal{H}_R
\end{equation}
in which $\mathcal{H}_S$, $\mathcal{H}_B$, and $\mathcal{H}_R$ represent the stretching, bending, and repulsive contributions, respectively. We treat each network segment as a Hookean spring with spring constant $\mu$, such that
\begin{equation}
\mathcal{H}_S = \frac{\mu}{2}\sum_{ ij }{\frac{\left(\ell_{ij}-\ell_{ij,0}\right)^2}{\ell_{ij,0}}}
\end{equation}
in which $\ell_{ij}$ and $\ell_{ij,0}$ are the length and rest length, respectively, of the segment connecting nodes $i$ and $j$. We add harmonic bond-bending interactions with energy scale $\kappa$ between connected segments as
\begin{equation}
\mathcal{H}_B = \frac{\kappa}{2}\sum_{ ijk}{\frac{\left(\theta_{ijk}-\theta_{ijk,0}\right)^2}{\ell_{ijk,0}}}
\end{equation}
in which $\theta_{ijk}$ and $\theta_{ijk,0}$ are the angle and rest angle between neighboring segments $ij$ and $jk$, $\ell_{ijk,0} = (\ell_{ij,0} + \ell_{jk,0})/2$, and the sum is taken over all connected network node triplets. Rest angles and rest lengths are defined such that the initial network structure corresponds to the zero energy configuration. We also include purely repulsive harmonic  interactions with energy scale $\mu_R$ as $\mathcal{H}_R = \mathcal{H}_{R,a} + \mathcal{H}_{R,b} $, where $\mathcal{H}_{R,a} $ resists overlap between pairs of inclusions and $\mathcal{H}_{R,b}$  resists overlap between inclusions and network nodes. The first is defined as
\begin{equation}
\mathcal{H}_{R,a} = \frac{\mu_R}{2}\sum_{mn}{\left(1-\frac{u_{mn}}{r_{mn}}\right)^2\Theta\left(1-\frac{u_{mn}}{r_{mn}}\right)}
\end{equation}
in which $u_{mn} = |\bm{u}_m - \bm{u}_n|$ is the distance between the central nodes of inclusions $m$ and $n$, $r_{mn}=r_m+r_n$ is the sum of their radii, and $\Theta$ is the Heaviside step function. The repulsive contribution between inclusions and network nodes is
\begin{equation}
\mathcal{H}_{R,b} = \frac{\mu_R}{2}\sum_{m}\sum_{i}{\left(1-\frac{u_{mi}}{r_{m}}\right)^2\Theta\left(1-\frac{u_{mi}}{r_{m}}\right)}
\end{equation}
in which $u_{mi} = |\bm{u}_m - \bm{u}_i|$ is the distance between the center node of inclusion $m$ and network node $i$ and the sums are taken over all inclusions $m$ and network nodes $i$.

\subsection*{Rheology simulation}
	 We consider quasistatic uniaxial strain $\varepsilon$ followed by simple shear strain $\gamma$,  applied relative to the initial reference configuration using generalized Lees-Edwards boundary conditions \cite{Lees1972}. In 3D, the deformation gradient tensor is
	 \begin{equation}
	 \bm{\Lambda}(\gamma,\varepsilon)=\begin{pmatrix} 1 & 0 & \gamma \\ 0 & 1 & 0 \\ 0 & 0 & 1 \end{pmatrix} \begin{pmatrix} 1 & 0 & 0\\ 0 & 1 & 0 \\ 0 & 0 & 1 + \varepsilon \end{pmatrix}.
	 \end{equation}
	 At each applied strain step, we numerically minimize $\mathcal{H}$ using the L-BFGS algorithm \cite{Nocedal2006}. About a given relaxed configuration at uniaxial strain $\varepsilon$, we compute the shear stress $\sigma_{xz}(\varepsilon)$ as
	 \begin{equation}
	 \sigma_{xz} = \frac{1}{V}\frac{\partial\mathcal{H}}{\partial\gamma}
	 \end{equation}
	in which $V$ is the volume of the deformed simulation box.
We apply small uniaxial strain steps of magnitude $|d\varepsilon| = 0.01$. At a given uniaxial strain, we apply small symmetric positive and negative shear strain steps  $d\gamma = 0.01$ to compute the apparent shear modulus
\begin{equation}
 G(\varepsilon) = \frac{\partial\sigma_{xz}(\varepsilon)}{\partial\gamma}
 \end{equation} 
  All data reported in this paper correspond to an average over 15 independently generated network samples.

\acknowledgements{This work was supported in part by the National Science Foundation Division of Materials Research (Grant DMR1826623) and the National Science Foundation Center for Theoretical Biological Physics (Grant PHY-1427654). J.L.S. acknowledges additional support from the Ken Kennedy Institute Graduate Fellowship and the Riki Kobayashi Fellowship in Chemical Engineering.}


%

\end{document}